\documentclass[floatfix,prl,twocolumn,letterpaper,lengthcheck,superscriptaddress,showpacs,amssymb,amsmath,amsfonts,aps,altaffilletter,nopreprintnumbers,showpacs,longbibliography]{revtex4-2}

\usepackage{xcolor}
\usepackage{graphicx}
\usepackage{dcolumn}
\usepackage{bm}
\usepackage{hyperref}

\def\massDomOnly{\ensuremath{84^{+13}_{-27}\,M_{\odot}}}
\def\massOvertone{\ensuremath{71^{+30}_{-16}\,M_{\odot}}}
\def\spinDomOnly{\ensuremath{0.84^{+0.09}_{-0.55}}}
\def\spinOvertone{\ensuremath{0.70^{+0.25}_{-0.47}}}

\def\tpeak{\ensuremath{t_{c}}}

\begin{document}

\title{Low evidence for ringdown overtone in GW150914 when marginalizing over time and sky location uncertainty}

\author{Alex Correia}
 \affiliation{Physics Department, University of Massachusetts Dartmouth, North Dartmouth, Massachusetts 02747, USA}
\author{Yi-Fan Wang}%
\affiliation{Max-Planck-Institut f{\"u}r Gravitationsphysik (Albert-Einstein-Institut), Am M{\"u}hlenberg 1, D-14476 Potsdam, Germany}
\author{Julian Westerweck}%
\affiliation{Max-Planck-Institut f{\"u}r Gravitationsphysik (Albert-Einstein-Institut), Callinstra{\ss}e 38, D-30167 Hannover, Germany}
\affiliation{Leibniz Universit{\"a}t Hannover, D-30167 Hannover, Germany}
\affiliation{Institute for Gravitational Wave Astronomy and School of Physics and Astronomy, University of Birmingham, Edgbaston, Birmingham B15 2TT, England, United Kingdom}
\author{Collin D. Capano}%
 \email{cdcapano@syr.edu}
\affiliation{Department of Physics, Syracuse University, Syracuse, New York 13244, USA}
\affiliation{Physics Department, University of Massachusetts Dartmouth, North Dartmouth, MA 02747, USA}
\date{\today}

\begin{abstract}
Tests of the no-hair theorem using astrophysical black holes involve the detection of at least two quasinormal modes (QNMs) in the gravitational wave emitted by a perturbed black hole. A detection of two modes---the dominant, $(\ell, m, n) = (2,2,0)$, mode and its first overtone, the $(2,2,1)$ mode---in the postmerger signal of the binary black hole merger GW150914 was claimed by Isi \textit{et al.}~\cite{isi-221-initial}, with further evidence provided by Isi \& Farr~\cite{isi-221-rebuttal}. However, Cotesta \textit{et al.}~\cite{cotesta-221} disputed this claim, finding that evidence for the overtone only appeared if the signal was analyzed before merger, when a QNM description of the signal is not valid. Because of technical challenges, both of these analyses fixed the merger time and sky location of GW150914 when estimating the evidence for the overtone. At least some of the contention can be attributed to fixing these parameters. Here, we surmount this difficulty and fully marginalize over merger time and sky-location uncertainty while doing a postmerger QNM analysis of GW150914.
We find that marginalizing over all parameters yields low evidence for the presence of the overtone, with a Bayes factor of $1.10\pm 0.03$ in favor of a QNM model with the overtone versus one without. The arrival time uncertainty of GW150914 is too large to definitively claim detection of the $(2,2,1)$ mode.
\end{abstract}

\maketitle

\section{Introduction}

The direct detection of gravitational waves has made it possible to perform ``black hole spectroscopy'' on real astrophysical black holes. Black holes that are formed from a binary black hole (BBH) merger are initially perturbed. The perturbation is radiated away in a gravitational-wave ``ringdown'' as the black hole settles down to a Kerr spacetime. The ringdown is a superposition of damped sinusoids, or quasinormal modes (QNMs), that are uniquely determined by three indices: two angular indices $\ell \geq 2$ and $m \leq |\ell|$, and an overtone number $n \geq 0$~\cite{Vishveshwara:1970zz,ringdown}. As a consequence of the no-hair theorem, we expect that the frequencies and damping times of all the QNMs are fully characterized by the final black hole's mass and spin. The Kerr nature of the final object can be tested if more than one QNM can be observed in a postmerger gravitational wave (GW)~\cite{no-hair-spec}. Should the QNMs be found to need more than two parameters to characterize their frequency and damping times it would indicate that the final object is not a Kerr black hole as expected from General Relativity (GR).

Black hole spectroscopy is challenging to accomplish in practice. The QNMs are exponentially damped, and so the signal-to-noise ratio (SNR) of the signal falls off quickly after merger. This would favor searching for QNMs immediately after merger, but it was thought that one had to wait $\sim10M$ (where $M$ is the mass of the black hole, in seconds) or longer (depending on the sensitivity of the detectors~\cite{London:2018gaq,Carullo:2018sfu}) after the time at which the GW amplitude peaks (\tpeak) for nonlinearities to radiate away and a QNM description of the signal to be valid~\cite{Buonanno:2006ui,Berti:2007fi,Kamaretsos:2011um,London:2014cma,Gupta:2018znn}. The $10M$ rule was arrived at by studying the stability of the ``fundamental'' $n=0$ modes in numerical signals. However, Giesler \textit{et al.}~\cite{giesler-overtone} show that including overtones of the dominant, $(\ell, m, n) = (2,2,0)$ mode---i.e., modes with $(2, 2, n \geq 1)$---yielded surprisingly good fits to a numerical relativity simulation at \tpeak{}. This led Giesler \textit{et al.}\ to claim that the spacetime could be described by a linearly perturbed black hole \emph{at the merger}. Isi \textit{et al.}~\cite{isi-221-initial} then performed an overtone analysis of the first detected binary black hole merger, GW150914~\cite{GW150914-observation}, and claimed detection of the $(2,2,1)$ mode at $3.6\,\sigma$. They performed a no-hair test using the overtone by allowing its frequency and damping time to vary in a Bayesian analysis, finding that it was broadly consistent with GR.

The validity of using a QNM description of the signal at \tpeak{} is controversial. Although it was known prior to the work of Giesler \textit{et al.} that overtones yielded a better fit to the GW at merger ~\cite{Buonanno:2006ui}, it was not clear whether this indicated that the spacetime was truly described by a linearly perturbed black hole (as claimed in Giesler \textit{et al.}) or if it was simply due to the overtones overfitting the signal (meaning that any sufficiently compatible signal could be fit by them). While Giesler \textit{et al.}~provided some evidence for the former, subsequent analyses have challenged their interpretation~\cite{Bhagwat:2019dtm,Jaramillo:2020tuu,Ma:2022wpv,Cheung:2022rbm,Mitman:2022qdl,Nee:2023osy,Baibhav:2023clw}, leaving the question unresolved.

Even the claim of a detection of an overtone in GW150914 by Isi \textit{et al.}~\cite{isi-221-initial} has been challenged. Cotesta \textit{et al.}~\cite{cotesta-221} performed an overtone analysis at multiple times around the purported \tpeak{} of GW150914. At each time they calculated the Bayes factor for a signal model consisting of the $(2,2,0){+}(2,2,1)$ QNMs versus one consisting of just the $(2,2,0)$ mode. Assuming equal prior weight for the two models, the Bayes factor $\mathcal{B} = Z_{220{+}221}/Z_{220}$ (where $Z_x$ is the Bayesian evidence for model $x$) gives the odds ratio that the overtone model is a better description of the signal. Cotesta \textit{et al.}~found that the Bayes factor was $\sim 1$ at \tpeak{}, indicating that the data were uninformative as to the presence of the $(2,2,1)$ mode. They claimed that positive evidence for the overtone only appeared if the QNM model was used \emph{prior} to the merger. Since a QNM model is not valid there, this would indicate that the overtone model was simply overfitting the data. The low Bayes factor was also consistent with what had been reported by Bustillo \textit{et al.}~\cite{CalderonBustillo:2020rmh} and in publications from the LIGO and Virgo Collaborations (LVC)~\cite{lvk-bayes}\footnote{It should be noted that the pipeline~\cite{pyring-methods, pyring-repo} used to report overtone Bayes factors by the LVC is the same as used in Cotesta \textit{et al.}, so the agreement between the two is not entirely surprising.}. However, in a rebuttal, Isi and Farr~\cite{isi-221-rebuttal} found that the Bayes factor for the overtone was nearly $100$ at \tpeak{}, reinforcing the initial finding by Isi \textit{et al.}~\cite{isi-221-initial}.

The disagreement between Isi \textit{et al.}\ and Cotesta \textit{et al.}\ may be due to several factors, including data conditioning choices and the manner in which the likelihood function is evaluated. The effects of these choices were investigated by Wang \textit{et al.}~\cite{Wang:2023xsy}. It is evident, however, that the choice of start time for the ringdown is one of the most important factors---if not the defining factor~\cite{Carullo:2023gtf}---in determining if the overtone is present. While they may disagree on the magnitude of the Bayes factor at a given time, both Cotesta \textit{et al.}\ and Isi and Farr found that its value changes rapidly as the ringdown start time is varied around \tpeak{}, increasing by several orders of magnitude in less than $5M$. For GW150914, $5M$ corresponds to $\sim 1.7\,$ms~\cite{GW150914-observation}. This is strikingly small; for comparison, the statistical uncertainty on \tpeak{} when using an inspiral-merger-ringdown (IMR) template to model the full signal is {$\sim 1.2\,$ms}~\cite{LIGOScientific:2016vlm,4ogc}

The statistical uncertainty in $\tpeak$ arises from an overall uncertainty in the arrival time of the signal in a single detector, and the uncertainty in the sky location of GW150914. The latter affects the relative offset in arrival times between the Hanford and Livingston detectors, which  observed the event~\cite{GW150914-observation}. A full accounting of the evidence for or against the overtone should marginalize over the time and sky location uncertainty of GW150914. However, the analysis pipelines used by Isi \textit{et al.}\ and Cotesta \textit{et al.}\ fix the arrival time and sky location of the signal to their maximum likelihood values from IMR studies.

Even when changing the start time to recalculate the Bayes factor, the sky location was still fixed in the previous analyses. This was done to keep the ringdown start time unchanged during these pipelines' Bayesian inference analysis.
Fixing the start time is necessary due to technical hurdles, discussed below, when calculating the likelihood. For this reason, almost every ringdown analysis that has been performed to date---both on GW150914 and other events---has fixed the sky location and merger time~\cite{lvk-bayes,isi-221-initial,cotesta-221,Capano:2021etf,isi-area-thrm,kastha-area-thrm,isi-221-rebuttal,Capano:2022zqm,Abedi:2023kot,Wang:2023xsy}.

Finch and Moore were able to marginalize over the sky location and arrival time in their postmerger analysis of GW150914~\cite{finch-moore}. They accomplished this by attaching an arbitrary signal model to the start of the ringdown. This had the effect of smoothly tapering the signal model to zero, thereby allowing them to use the traditional frequency-domain likelihood function that is used in standard parameter estimation analyses. Doing so, they recovered Bayes factors $B_{220}^{221}$ between 0.1 and 10, depending on the prior chosen on $t_c$. Their prior choices consisted of Gaussian distributions of width $1M$ centered at various $t_c$ values.

By using the traditional frequency-domain likelihood function, Finch and Moore coupled information from the premerger part of the signal with the postmerger QNM model. The frequency-domain likelihood is equivalent to the convolution of the ``whitened'' data with the whitened signal model. The whitening filter couples adjacent points in time over several milliseconds. This means, for example, that the beginning of the whitened ringdown will be a weighted sum over a time span starting before the merger that is on order with the damping time of the overtone ($\sim 1.4\,$ms). Consequently, the recovered QNM parameters will be influenced by both the choice of signal model and the data just prior to the start of the ringdown.

Here, we overcome the technical hurdles that arise when marginalizing over time and sky location uncertainty while doing a purely QNM analysis of the postmerger portion of GW150914. We simultaneously analyze the premerger data using the inspiral-merger portion of an IMR template waveform. The pre and postmerger portions of the analysis only share a common sky location and geocentric $\tpeak$. This allows us to use the full signal to isolate the binary, giving the most accurate assessment of the timing uncertainty, while still maintaining independence between the pre and postmerger analyses.
In a companion paper~\cite{Correia:2023ipz} we show that it is necessary to model the entire observable signal in order to marginalize over timing uncertainties. Consequently, our analysis here is the most complete accounting of all uncertainties that can be preformed when using a QNM-only model of the postmerger signal.

\begin{figure*}[t]
    \centering
    \includegraphics[width=\textwidth]{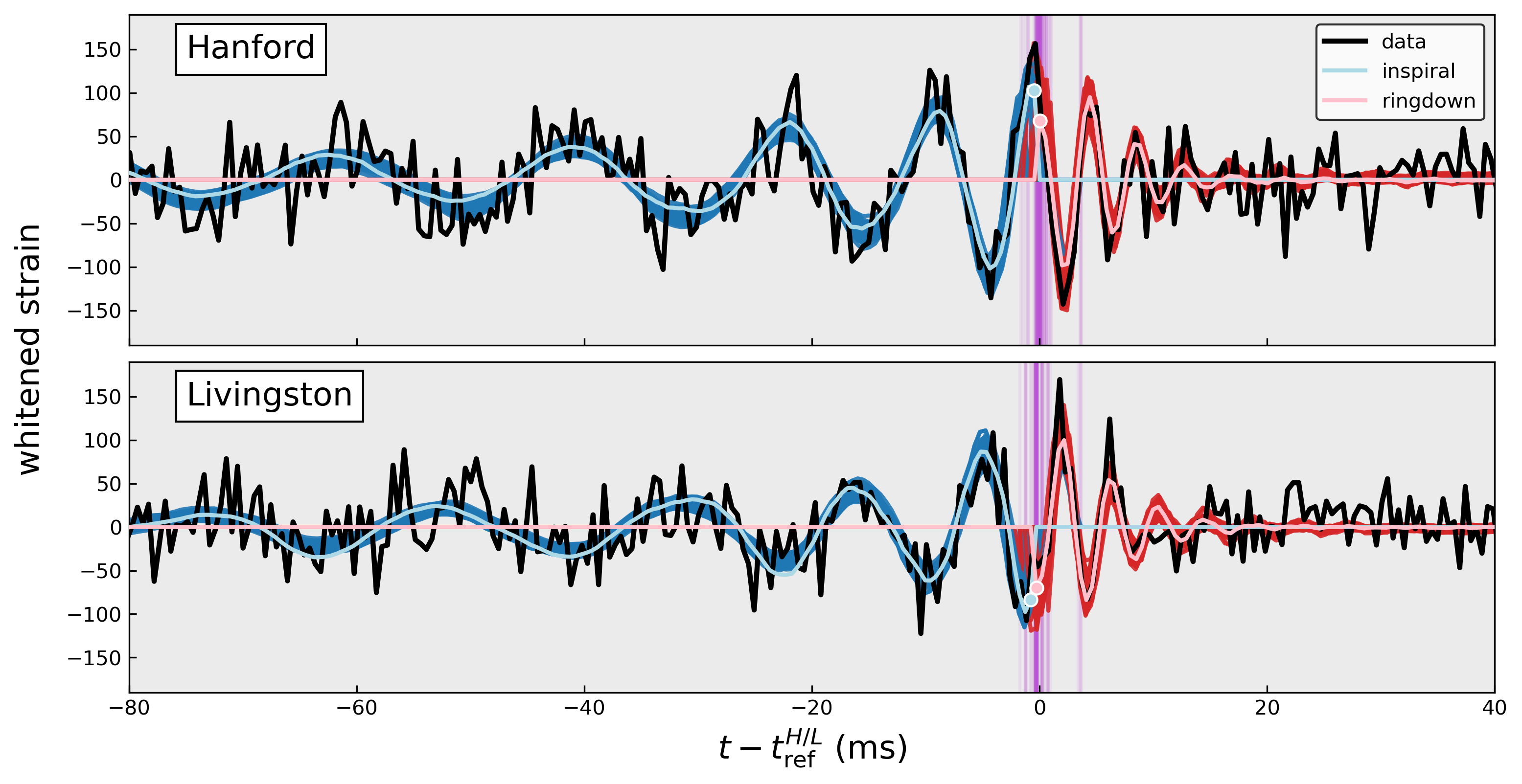}
    \caption{Whitened strain in the Hanford (top) and Livingston (bottom) detectors around GW150914. Plotted is the original data (black line) along with the maximum likelihood waveform, which consists of an inspiral portion (blue lines) and ringdown portion (red lines). The inspiral and ringdown models are independent of each other, and only share a common sky location and geocentric $t_{c}$. Ringdown (inspiral) waveforms are gated and in-painted before (after) the $t_{c}$ in each detector; the first (last) nonzero point in the waveform is indicated by the solid dot. The data are also gated and in-painted before (after) the detector $t_c$ for the ringdown (inspiral) model, so that neither model simultaneously includes the same data point from the signal in the likelihood. Individual waveforms drawn from the upper $90$th percentile of the posterior are also plotted to indicate the uncertainty region. Vertical purple lines indicate the gate time for each of the waveforms plotted. The time axis is zeroed with respect to the maximum likelihood $t_c$ in each detector $t_{\rm ref}^{H/L}$ from the 4-OGC IMR analysis~\cite{4ogc}.}
    \label{fig:waveforms}
\end{figure*}

\section{Methods}

An in-depth description of our analysis is presented in a following paper~\cite{Correia:2023ipz}. Here, we give a brief overview.

The Bayesian evidence for a signal model is given by marginalizing the likelihood function times the prior over all the model parameters $\boldsymbol{\vartheta}$. In GW astronomy it is common to assume that each interferometer outputs independent, wide-sense stationary Gaussian noise with zero mean in the absence of a GW signal. Under this assumption, the likelihood function is given by
\begin{equation}
\label{eqn:likelihood}
p(\mathbf{s}|\boldsymbol{\vartheta}, h) = \prod_{D=H, L,...} \frac{\exp[-\frac{1}{2} \mathbf{r}_D^\mathsf{T} \boldsymbol{\Sigma}_D^{-1} \mathbf{r}_D]}{\sqrt{(2\pi)^{N} \det \boldsymbol{\Sigma}_D}}.
\end{equation}
Here, the product is taken over the number of detectors, $\mathbf{r}_D = \mathbf{s}_D - \mathbf{h}_D(\boldsymbol{\vartheta})$ is the residual after subtracting the signal model $\mathbf{h}(\boldsymbol{\vartheta})$ from the time-domain data $\mathbf{s}_D$, which contains $N$ samples. The elements of the covariance matrix $\boldsymbol{\Sigma}_D$ are given by the autocorrelation function of the data.

In standard GW parameter estimation (in which $\mathbf{h}$ models the entire observable signal), $\boldsymbol{\Sigma}_D$ can be approximated as a circulant matrix. That yields an analytic solution for $\boldsymbol{\Sigma}_D^{-1}$, which in turn yields a weighted inner product for the log likelihood that is most efficiently calculated in the frequency domain. However, since a QNM template only models the ringdown portion of the signal, data prior to the start of the ringdown must be excised from the analysis, else biases will arise. Excising data breaks the circulant approximation of $\boldsymbol{\Sigma}_D$, and so the standard frequency-domain likelihood function cannot be used.

There are two approaches around this. One is to numerically invert $\boldsymbol{\Sigma}_D$ and explicitly calculate the matrix multiplication in Eq.~\eqref{eqn:likelihood}. This is what is done by the ``\textsc{ringdown}''~\cite{isi-ringdown} and ``\textsc{pyring}''~\cite{pyring-methods, pyring-repo} analyses used by Isi \textit{et al.}\ and Cotesta \textit{et al.}, respectively. Numerically inverting $\boldsymbol{\Sigma}_D$ is computationally expensive but needs only to be done once per analysis if the start time of the ringdown does not change. If the start time of the ringdown does change (meaning different times need to be excised from the data), then $\boldsymbol{\Sigma}^{-1}_D$ and $\det \boldsymbol{\Sigma}_D$ need to be recalculated, making the analysis substantially more expensive. 

The other approach is to use ``gating and in-painting'' to solve the likelihood~\cite{gating-inpainting}. This is mathematically equivalent to the brute force numerical method, but is more efficient for longer stretches of data~\cite{Isi:2021iql,Wang:2023xsy}. In this approach, a gate is applied to a stretch of data, within which the data are zeroed out. The gated portion is then in-painted so that it contributes nothing to the likelihood. For a ringdown-only analysis the gate ends at the start of the ringdown; we use a $1\,$s gate to ensure that the rest of the observable signal is removed. The gate shifts around in time as the start time of the ringdown is varied, thereby removing more or less of the signal from the analysis.

The cost of computing the numerator in Eq.~\eqref{eqn:likelihood} with the gating and in-painting method is the same regardless of the start time of the ringdown. However, $\det \boldsymbol{\Sigma}_D$ still needs to be calculated, as its value will change for different ringdown start times. For this reason, studies that have used gating and in-painting have still fixed the start time and sky location when doing QNM analyses~\cite{Capano:2021etf,kastha-area-thrm,Capano:2022zqm,Siegel:2023lxl,Abedi:2023kot,Wang:2023xsy}.

In Ref.~\cite{Correia:2023ipz}, we show that to high precision $\det \boldsymbol{\Sigma}_D$ can be approximated by a simple linear interpolation. This allows us to quickly evaluate all terms in Eq.~\eqref{eqn:likelihood} using gating and in-painting, regardless of the ringdown start time, thereby allowing for time and sky-location marginalization.

We also show in Ref.~\cite{Correia:2023ipz} that it is not possible to only model part of a signal when marginalizing over $t_c$ and sky location. Marginalizing over these parameters while only modeling part of the signal means that the portion of data that is excised from the analysis is allowed to shift around in time. If the shift is such that the entire observable signal is excised from the analysis, then a relatively large likelihood can be obtained for any parameter value by reducing the overall amplitude of the template to noise level. Consequently, the posterior probability will favor excising as much of the signal as possible, even the part that the template is meant to model. This means that for a ringdown-only analysis the start time of the template will be pushed to as \emph{late} as the prior will allow. Likewise, if an inspiral-only analysis is performed (in which the postmerger is excised and the premerger is modeled with a template) then the end time of the template will be pushed to as \emph{early} as the prior will allow. It is necessary to have a signal model for all times that a signal is observable in the data when doing any sort of marginalization over arrival time.

We overcome this problem by simultaneously yet independently modeling the pre and postmerger data. We treat the pre and postmerger signal as separate analyses that share a common set of hyperparameters---the sky location and geocentric $\tpeak$. We use gating and in-painting for both. For the premerger analysis, we excise the data \emph{after} $\tpeak$ while for the postmerger analysis, we excise the data \emph{before} \tpeak{}. The premerger portion of the signal is modeled using the \textsc{IMRPhenomXPHM} approximant~\cite{imrphenomxphm}, which we zero out for all times after $\tpeak$. The postmerger signal is modeled using a QNM model consisting of the $(2,2,0){+}(2,2,1)$ modes, which is zeroed out for all times before $\tpeak$. The final mass and final spin in the QNM model are varied over uniform priors such that $M_f \in [10, 200]M_{\odot}$ and $\chi_f \in [-0.99, 0.99]$. For the coalescence time we use a uniform prior spanning $t_{\rm ref} \pm 0.1\,$s, where $t_{\rm ref}$ is the maximum likelihood geocentric coalescence time from Ref.~\cite{4ogc}. This is the same prior width on $t_c$ as used in standard IMR analyses.

For the $(2,2,0)$ amplitude $A_{220}$ we sample over the range $[10^{-25}, 8\times10^{-17}]$ uniformly in log base 10. For the $(2,2,1)$ amplitude $A_{221}$ we use a uniform prior on $A_{221}/A_{220} \in [0, 5)$. We additionally impose the constraint that the $(2,2,0)$ mode's contribution to the postmerger SNR be $>2$. This is to ensure that at least the fundamental mode is present in the postmerger signal. Otherwise, we obtain samples where the $(2,2,1)$ mode appears to have large amplitude, but is in fact matching the $(2,2,0)$ mode in the signal (the ``label switching'' problem).

Gating the pre and postmerger means that no coupling occurs over the boundary due to whitening. Simultaneously modeling both portions has the added benefit that we use the entire signal to constrain the sky location. We use \textsc{PyCBC Inference}~\cite{Biwer:2018osg} with the \textsc{dynesty} sampler~\cite{speagle:2019} to perform the Bayesian analysis.

\section{Results and Discussion}

\begin{figure}
    \centering
    \includegraphics[width=\columnwidth]{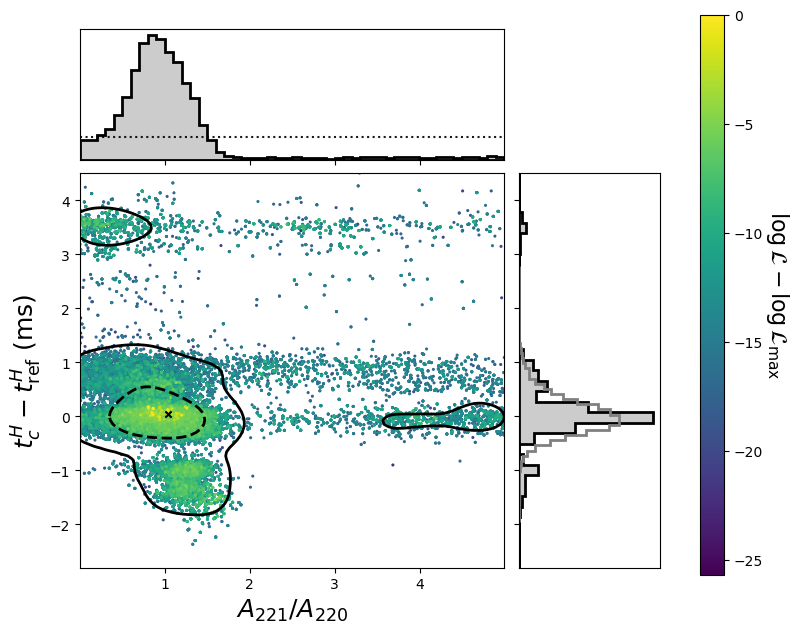}
    \caption{Marginal posteriors on $A_{221}/A_{220}$ and the coalescence time as measured in the Hanford detector $t_c^H$; similar results are obtained for the Livingston detector (not shown). The horizontal axis shows the ratio between the $(2,2,1)$ and $(2,2,0)$ mode amplitudes. The vertical axis shows the difference between the measured $t^H_c$ and the reference time $t^H_{\rm ref}$, which is the maximum likelihood $t^H_c$ from Ref.~\cite{4ogc}. The filled marginal histograms represent the one-dimensional distributions for each parameter. The prior on $A_{221}/A_{220}$ is overlaid on the corresponding marginal posterior with a dotted black line. The 4-OGC $t^H_c$ posterior is overlaid on the $t_c$ marginal posterior in gray. Each point in the two-dimensional scatter plot is colored according to its log likelihood value relative to the maximum log likelihood. The maximum likelihood point is denoted with a black cross. The two-dimensional credible regions are indicated by the contour lines; shown are the 50th (dashed contour) and 90th (solid contours) percentiles. The Bayes factor for the $(2,2,1)$ mode is estimated from the $A_{221}/A_{220}$ marginal posterior using the Savagey-Dickey ratio. This is the ratio between the posterior density (solid black line) and the prior density (dotted black line) at $A_{221}/A_{220}=0$. }
    \label{fig:result}
\end{figure}

The results of our analysis are summarized in Figs.~\ref{fig:waveforms} and~\ref{fig:result}. Figure~\ref{fig:waveforms} shows the recovered inspiral-only and ringdown-only templates. In Fig.~\ref{fig:result} we plot the marginal posterior on the $(2,2,1)$ amplitude $A_{221}$ and $\tpeak$ in the Hanford detector. The median and $90\%$ credible interval for the $(2,2,1)$ amplitude is $A_{221}/A_{220} = 0.9_{-0.7}^{+1.8}$; for the coalescence time it is $t^H_c - t^H_{\rm ref} = 0.0_{-1.1}^{+1.0}\,\text{ms}$ in Hanford and $t^L_c - t^L_{\rm ref} = 0.2_{-1.0}^{+1.1}\,\text{ms}$ in Livingston.\footnote{We use the maximum likelihood sky location and coalesence time from the 4-OGC analysis~\cite{4ogc} to calculate the reference GPS time in each detector. Specifically, we use $t^H_{\rm ref} = 1126259462.423701$ and $t^L_{\rm ref} = 1126259462.416674$.}

\begin{figure}
    \centering
    \includegraphics[width=0.45\textwidth]{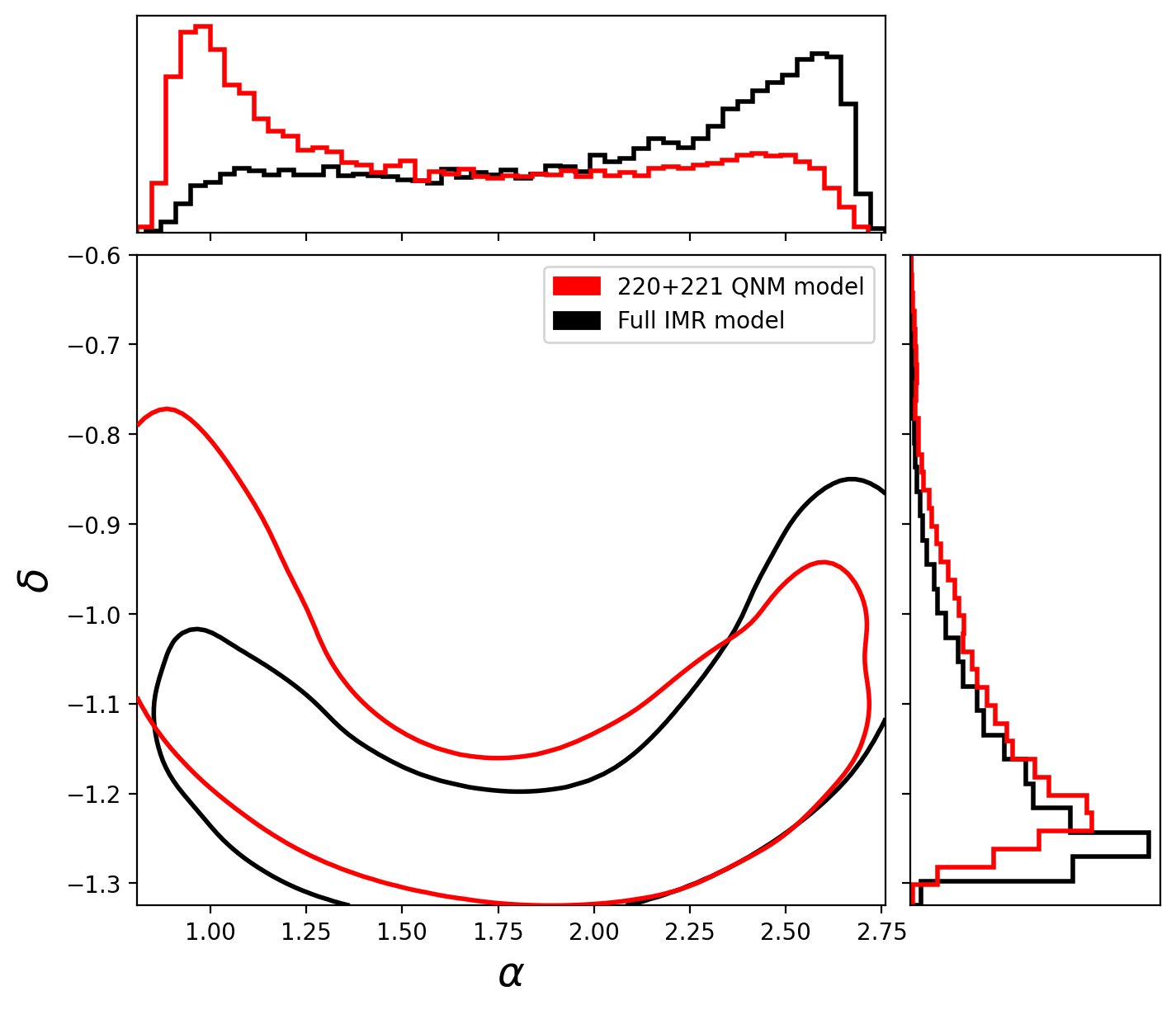}
    \caption{Sky location posteriors for our analysis depicted in Fig.~\ref{fig:result} and a full IMR analysis from 4-OGC \cite{4ogc}. Contours indicate the 90th percentile of points for each run. The red contour denotes the analysis conducted with sky and time marginalization described in Methods. The black contour denotes the 4-OGC analysis.}
    \label{fig:result_sky}
\end{figure}

Figure \ref{fig:result_sky} compares the sky location posteriors of our analysis with a full IMR analysis from the fourth Open Gravitational-wave Catalog (4-OGC) \cite{4ogc}. Our analysis recovers most of the 90th percentile of the 4-OGC posterior. Our marginal posterior on coalescence time is also consistent with the full IMR analysis (see Fig.~\ref{fig:result}). This indicates that our sky and time marginalization implementation is able to recover nominal sky parameters for GW150914. 

\begin{figure}
    \centering
    \includegraphics[width=0.45\textwidth]{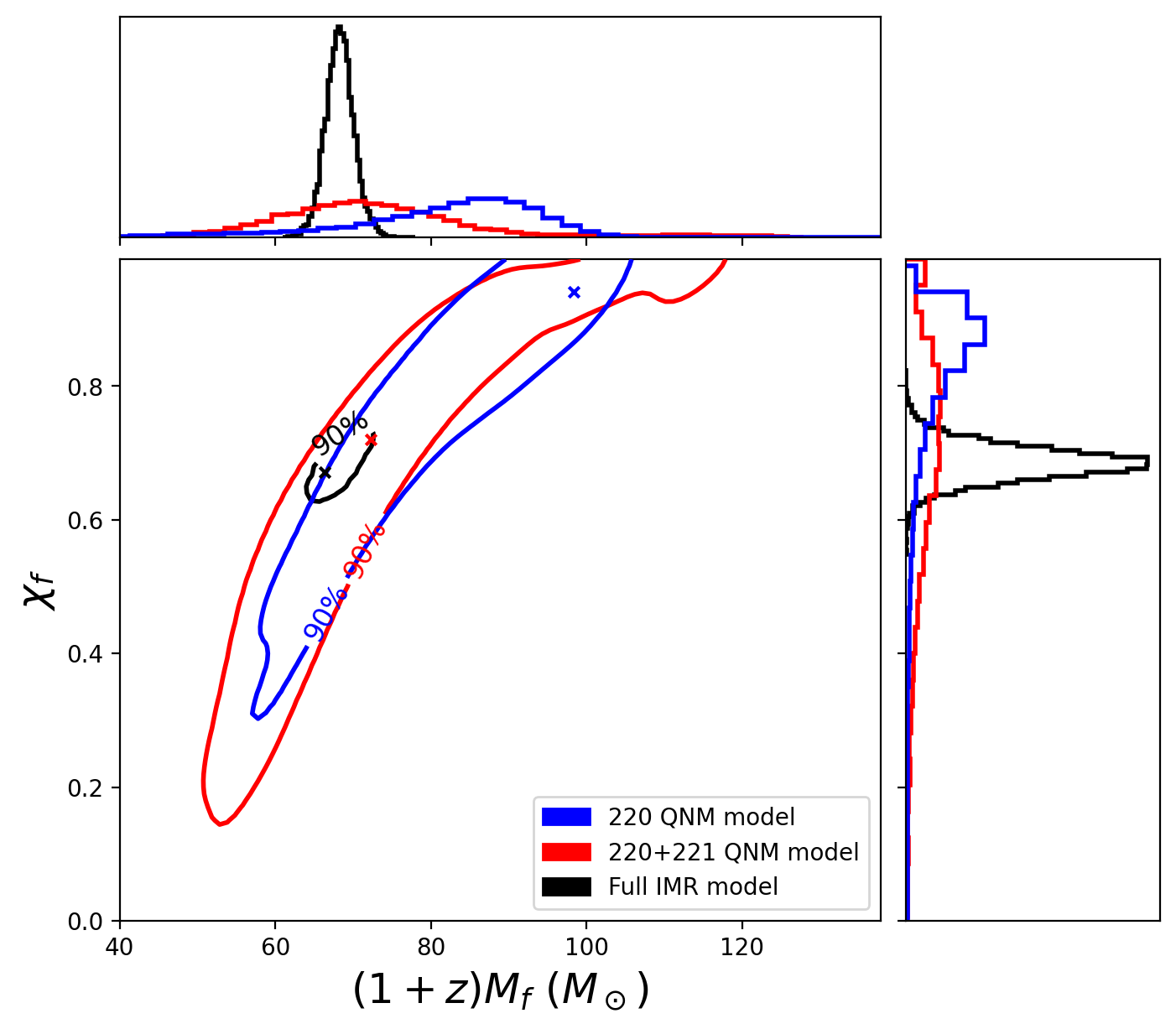}
    \caption{Final mass and spin posteriors for our analysis depicted in Fig.~\ref{fig:result}, a $(2,2,0)$-only ringdown analysis, and a full IMR analysis from 4-OGC \cite{4ogc}. Contours indicate the 90th percentile of points for each run. The red contour denotes our ringdown analysis with both the $(2,2,0)$ and $(2,2,1)$ modes. The blue contour denotes an analysis with only the dominant mode present in the ringdown. The black contour denotes the 4-OGC analysis. The maximum likelihood points of the models are indicated by the crosses. The median and 90\% credible intervals from the overtone model are \massOvertone{} and \spinOvertone{} for the red-shifted mass and spin, respectively, while for the $(2,2,0)$-only model they are \massDomOnly{} and \spinDomOnly{}.}
    \label{fig:result_mass_spin}
\end{figure}

We estimate the Bayes factor $\mathcal{B}$ between $Z_{220+221}$ and $Z_{220}$ by calculating the Savage-Dickey ratio \cite{savage-dickey} at $A_{221}/A_{220} = 0$. Doing so, we obtain $\mathcal{B} = 1.10 \pm 0.03$, indicating low support in favor of the $(2,2,1)$ mode. Similarly, the median of the posterior distribution for $A_{221}$ deviates from 0 by $1.2 \sigma$ \footnote{The $1.2\sigma$ quoted here was computed using the relative amplitude posterior $A_{221}/A_{220}$. Recalculating using the absolute $A_{221}$ posterior yields the same result.}. The reason for the small Bayes factor for the $(2,2,1)$ mode is evident from the marginal posterior plot in Fig.~\ref{fig:result}. While the maximum likelihood point is at $A_{221}/A_{220} = 1.03$ and  $t^H_c - t^H_{\rm ref} = 0.03\,$ms, the $A_{221}/A_{220}$ posterior has support at zero over much of the $t^H_c$ posterior (similarly for $t^L_c$). It is only at earlier times that the posterior on $A_{221}$ is markedly shifted away from zero. This is consistent with previous studies which found the evidence for the $(2,2,1)$ mode increased at earlier times~\cite{cotesta-221,isi-221-rebuttal,Wang:2023xsy}. However, due to the relatively large timing uncertainty---our credible interval on $t^{H/L}_c$ spans roughly $5M$---we cannot confidently claim detection of the overtone, as reflected by the Bayes factor. A signal with more SNR in the ringdown is needed.

Including the overtone does yield a mass and spin estimate of the final black hole that is more consistent with the full IMR analysis. This can be seen in Fig.~\ref{fig:result_mass_spin}, which compares the redshifted final mass and final spin parameters of the full IMR analysis to the QNM model that includes the (2,2,1) overtone, and to the QNM model that includes only the (2,2,0) mode \footnote{To obtain the $(2,2,0)$-only posterior on mass and spin, we preform a full Bayesian analysis in which the $(2,2,0)$-only QNM starts at $t_c$.}. This is consistent with previous analyses of GW150914~\cite{LIGOScientific:2016lio}, which also found that using a $(2,2,0)$-only QNM model at merger yielded biased results. Indeed, a major motivation for including the $(2,2,1)$ mode is that it yields better mass and spin estimates at merger~\cite{giesler-overtone}.

However, the overtone posterior is also \emph{wider}: the 90\% credible intervals from the overtone model span $46\,M_{\odot}$ and $0.72$ for the red-shifted mass and spin, respectively, while for the $(2,2,0)$-only model they span $40\,M_{\odot}$ and $0.64$. It is therefore unclear whether the better fit is simply due to the overtone model's additional freedom. This is borne out by the Bayes factor. A key feature of Bayes factors is they account for model complexity. That it is $\sim 1$ between the overtone and $(2,2,0)$-only models indicates that the better fit obtained by the former is offset by the Occam penalty from its larger number of parameters.

A detection of the $(3,3,0)$ subdominant QNM in the ringdown of GW190521 was claimed by Capano \textit{et al.}~\cite{Capano:2021etf}\footnote{See also the work by Siegel \textit{et al.}~\cite{Siegel:2023lxl}, which found evidence for the $(2,1,0)$ and $(3,2,0)$ modes in the same event.}. The sky location was also fixed in that analysis, and the Bayes factor was maximized over a grid of possible merger times. Our results here raise the question of whether the evidence for the subdominant mode in GW190521 would also decrease if the sky location and $t_c$ were marginalized over.

A reanalysis of GW190521 is beyond the scope of this paper. However, there are a few differences between the overtone claim in GW150914 and the $(3,3,0)$ claim in GW190521 that indicate that sky/time marginalization will not negatively affect the Bayes factor for the latter, and in fact may increase it. First, the SNR in the ringdown of GW190521 is about twice that of GW150914~\cite{Capano:2021etf,GW150914-observation}. Second, the difference in arrival times between Hanford and Livingston for GW190521 is $\sim 0.1\,\tau_{330}$~\cite{4ogc}, where $\tau_{330}$ is the damping time of the putative $(3,3,0)$ mode. For comparison, the difference in arrival times in units of the overtone damping time is $\sim 5\,\tau_{221}$ for GW150914. Marginalizing over sky location---which roughly corresponds to marginalizing over uncertainty in the difference in arrival times---is likely to have less of an effect on the evidence for the GW190521 $(3,3,0)$ mode.

Assuming the sky location used by Capano \textit{et al.}, the evidence for the $(3,3,0)$ mode stays elevated across a large range of times after merger~\cite{Capano:gw190521-data-release}. In contrast, the evidence for the $(2,2,1)$ mode for GW150914 drops quickly after merger. A naive marginalization over the discrete times published by Capano \textit{et al.}\---accomplished by summing the log evidence over those times---yields a Bayes factor for the $(3,3,0)$ mode of $\sim5000$.\footnote{We report the Bayes factor for the $(3,3,0)$ mode as being $Z_{220{+}330}/\max(Z_{220},Z_{220{+}221})$, as was done by Capano \textit{et al.}~\cite{Capano:2021etf}.} A reanalysis that uses a model for the pre and postmerger signal and marginalizes over time and sky-location uncertainty is necessary to fully evaluate the evidence for the $(3,3,0)$ mode, and any other subdominant QNM, in GW190521. Nevertheless, our cursory analysis indicates that the evidence for the $(3,3,0)$ mode in GW190521 may in fact increase with full sky and time marginalization. We plan to revisit this in a future study.

Returning to overtones, our results show that GW150914 is too weak to definitively claim detection of the $(2,2,1)$ mode. Gravitational-wave detectors are continuing to improve in sensitivity~\cite{LIGOScientific:2021djp,4ogc,LVKstatus}. A signal with sufficient SNR to resolve questions about the presence of one or more overtones is likely to eventually be observed.\footnote{Although larger SNR may result in an ambiguous number of overtones needed to fully model the merger, raising questions of overfitting~\cite{CalderonBustillo:2020rmh}.}
When such an event occurs, it will be necessary to fully marginalize over both time and sky-location uncertainties when performing a QNM analysis of the ringdown, as we have done here.

All computations were performed on Unity, a collaborative, multi-institutional high-performance computing cluster managed by UMass Amherst Research Computing and Data.

This research was conducted using \textsc{PyCBC}~\cite{pycbc-software}. Data are available in Ref.~\cite{data-release}.

\section{acknowledgments}

We wish to thank Gregorio Carullo and Juan Calderon Bustillo for providing helpful comments. A.C. was supported by funds from the Massachusetts Space Grant Consortium. C.C. acknowledges support from NSF Grant No.~PHY-2309356.

This research has made use of data or software obtained from the Gravitational Wave Open Science Center (\footnote{\href{gwosc.org}{gwosc.org}}), a service of the LIGO Scientific Collaboration, the Virgo Collaboration, and KAGRA. This material is based upon work supported by NSF's LIGO Laboratory which is a major facility fully funded by the National Science Foundation, as well as the Science and Technology Facilities Council (STFC) of the United Kingdom, the Max-Planck-Society (MPS), and the State of Niedersachsen/Germany for support of the construction of Advanced LIGO and construction and operation of the GEO600 detector. Additional support for Advanced LIGO was provided by the Australian Research Council. Virgo is funded, through the European Gravitational Observatory (EGO), by the French Centre National de Recherche Scientifique (CNRS), the Italian Istituto Nazionale di Fisica Nucleare (INFN) and the Dutch Nikhef, with contributions by institutions from Belgium, Germany, Greece, Hungary, Ireland, Japan, Monaco, Poland, Portugal, and Spain. KAGRA is supported by Ministry of Education, Culture, Sports, Science and Technology (MEXT), Japan Society for the Promotion of Science (JSPS) in Japan; National Research Foundation (NRF) and Ministry of Science and ICT (MSIT) in Korea; and Academia Sinica (AS) and National Science and Technology Council (NSTC) in Taiwan.

\bibliography{references}

\end{document}